\documentstyle[12pt]{article}

\advance \topmargin by -\headheight
\advance \topmargin by -\headsep

\evensidemargin \oddsidemargin
\def\ie{{\em i.e.}}

\def\ie{\hbox{\it i.e.}}

\def\CC{{\mathchoice
{\rm C\mkern-8mu\vrule height1.45ex depth-.05ex 
width.05em\mkern9mu\kern-.05em}
{\rm C\mkern-8mu\vrule height1.45ex depth-.05ex 
width.05em\mkern9mu\kern-.05em}
{\rm C\mkern-8mu\vrule height1ex depth-.07ex 
width.035em\mkern9mu\kern-.035em}
{\rm C\mkern-8mu\vrule height.65ex depth-.1ex 
width.025em\mkern8mu\kern-.025em}}}

\def\RR{{\rm I\kern-1.6pt {\rm R}}}

\def\ZZ{{\rm Z}\kern-3.8pt {\rm Z} \kern2pt}

\def\np{Nucl. Phys.}
\def\pl{Phys. Lett.}
\def\prl{Phys. Rev. Lett.}
\def\pr{Phys. Rev.}

\def\jhep{J. High Energy Phys.}

\newcommand{\beq}{\begin{equation}}
\newcommand{\eeq}{\end{equation}}
\newcommand{\rc}{\nonumber\\}
\newcommand{\bear}{\begin{eqnarray}}
\newcommand{\eear}{\end{eqnarray}}

\newfont{\namefont}{cmr10}
\newfont{\addfont}{cmti7 scaled 1440}
\newfont{\boldmathfont}{cmbx10}
\newfont{\headfontb}{cmbx10 scaled 1728}
\begin{document}
\begin{titlepage}

\begin{center} \Large \bf M-Theory Giant Gravitons with C field

\end{center}

\vskip 0.3truein
\begin{center} 
J. M. Camino
\footnote{e-mail:camino@fpaxp1.usc.es}
and 
A.V. Ramallo
\footnote{e-mail:alfonso@fpaxp1.usc.es}

\vspace{0.3in}

Departamento de F\'\i sica de
Part\'\i culas, \\ Universidad de Santiago\\
E-15706 Santiago de Compostela, Spain. 
\vspace{0.3in}

\end{center}
\vskip 1truein

\begin{center}
\bf ABSTRACT
\end{center} 

We find giant graviton configurations of an M5-brane probe in the D=11
supergravity background generated by a stack of non-threshold (M2,M5) bound
states. The M5-brane probe shares three directions with the background and
wraps a  two-sphere transverse to the bound states. For a particular value of
the worldvolume gauge field of the PST formalism, there exist solutions of the
equations of motion for which the M5-brane probe behaves as a wave propagating
in the (M2,M5) background. We have checked that the probe breaks the
supersymmetry of the background exactly as a massless particle moving along  the
trajectory of its center of mass.

\vskip4.5truecm
\leftline{US-FT-12/01\hfill October 2001}
\leftline{hep-th/0110096}
\smallskip
\end{titlepage}
\setcounter{footnote}{0}

\setcounter{equation}{0}
\section{Introduction}
\medskip
The so-called giant gravitons are configurations of branes which behave as an
expanded massless particle. They were introduced in ref. \cite{GST} for branes
moving in a spacetime of the type $AdS_m\times S^{p+2}$ and generalized in
refs. \cite{DTV}-\cite{gravitons} for more general near-horizon brane geometries.
The supersymmetry of the giant graviton configurations in $AdS_m\times S^{p+2}$
spacetimes was studied in refs. \cite{GMT, HHI}, where it was proved that they
preserve the same supersymmetries as the point-like graviton in the same
spacetime (see also \cite{DJM}-\cite{Kim}). 

The general mechanism underlying the construction of ref. \cite{GST} is the
coupling of the brane probe to the background gauge field. The flux of this
gauge field  captured by the wrapped brane probe stabilizes it against
shrinking, which allows the existence of stable solutions behaving as massless
particles. This situation was generalized in ref. \cite{gravitons}, where 
giant gravitons for a type II background created by a stack of
non-threshold  (D(p-2), Dp) bound states were found. In this case, the probe is a
D(8-p)-brane which wraps  the $S^{6-p}$ sphere transverse to the background, and
is extended along two directions parallel to it. This configuration is such that
the probe captures both the Ramond-Ramond flux and the flux of the Kalb-Ramond
$B$ field. 

In this paper we extend the analysis of ref. \cite{gravitons} to M-theory
backgrounds generated by a stack of non-threshold bound states of the type
(M2,M5). The corresponding solution of the D=11 supergravity equations was given
in ref. \cite{ILPT} and used as supergravity dual of a non-commutative field
theory in ref. \cite{NC}. As suggested in
\cite{gravitons}, the probe we will consider is an M5-brane wrapped on an $S^2$
transverse sphere and extended along three directions parallel to the
background. We will show that, after switching on a particular value of the
worldvolume gauge field, one can find giant graviton solutions of the
corresponding worldvolume equations of motion. We will also analyze the
supersymmetry of the problem, and we will show that our M5-brane configurations
break supersymmetry exactly in the same way as a wave which propagates with the
velocity of the center of mass of the M5-brane probe.

\setcounter{equation}{0}
\section{The supergravity background}
\medskip
The metric for the eleven dimensional supergravity solution of the background
we will consider is \cite{ILPT}:
\bear
ds^2&=&f^{-1/3}\,h^{-1/3}\,\Big[\,-(\,dx^0\,)^2\,+\,(\,dx^{1}\,)^2\,
+\,(\,dx^{2}\,)^2\,+\,h\, \Big((\,dx^{3}\,)^2\,
+\,(\,dx^{4}\,)^2\,+\,(\,dx^{5}\,)^2\Big) \,\Big]\,+\rc\rc
&&+\,f^{2/3}\,h^{-1/3}\, \Big[\,dr^2\,+\,r^2\,d\Omega_{4}^2\,\Big]\,\,,
\label{uno}
\eear
where $d\Omega_{4}^2$ is the line element of a unit 4-sphere and the functions
$f$ and $h$ are given by:
\bear
f&=&1\,+\,{R^{3}\over r^{3}}\,\,,\rc\rc
h^{-1}&=&\sin^2\varphi\,f^{-1}\,+\,\cos^2\varphi\,\,.
\label{dos}
\eear
The metric (\ref{uno}) is the one generated by a stack of parallel
non-threshold (M2,M5) bound states. The M5-brane component of this bound state
is extended along the directions $x^0,\cdots, x^5$, whereas the M2-brane lies
along $x^0,x^1,x^2$. The angle $\varphi$ in eq. (\ref{dos})
determines the mixing of the M2- and M5-branes in the bound state and the
``radius" $R$ is given by $R^{3}\cos\varphi\,=\,\pi\,N\,l_{p}^3$, where $l_p$ is
the Planck length in eleven dimensions and $N$ is the number  of bound states of
the stack. The solution of D=11 supergravity is also characterized \cite{ILPT} by
a non-vanishing value of the four-form field strength 
$F^{(4)}$, namely:
\bear
F^{(4)}\,&=&\,\sin\varphi\,\partial_r(f^{-1})\,dx^0\wedge dx^1\wedge
dx^2\wedge dr\,-\,3 R^3\cos\varphi \,\epsilon_{(4)}\,-\cr\cr
&&-\tan\varphi \,\partial_r(hf^{-1})\,dx^3\wedge dx^4\wedge
dx^5\wedge dr\,\,,
\label{tres}
\eear
where $\epsilon_{(4)}$ represents the volume form of the unit $S^4$. The field
strength $F^{(4)}$ can be represented as the exterior derivative of a
three-form  potential  $C^{(3)}$, \ie\ as $F^{(4)}\,=\,dC^{(3)}$. In order to
obtain the explicit form of $C^{(3)}$, let us introduce a particular set of
coordinates for the transverse $S^4$ \cite{GST}. Let $\rho$ and $\phi$ take
values in the range $0\le\rho\le 1$ and $0\le\phi\le 2\pi$ respectively. Then,
the line element $d\Omega_{4}^2$ can be written as:
\beq
d\Omega_{4}^2\,=\,{1\over 1-\rho^2}\,d\rho^2\,+\,
(1\,-\,\rho^2\,)\,d\phi^2\,+\,\rho^2\,d\Omega_{2}^2\,\,,
\label{cuatro}
\eeq
where $d\Omega_{2}^2$ is the metric of a unit $S^2$ (which we will parametrize
by means of two angles $\theta^1$ and $\theta^2$). In these coordinates one can
take $C^{(3)}$ as:
\bear
C^{(3)}\,&=&\,-\sin\varphi\,f^{-1}\,dx^0\wedge dx^1\wedge
dx^2\,-\, R^3\cos\varphi \,\rho^3\,d\phi\wedge\epsilon_{(2)}\,+\cr\cr
&&+\tan\varphi \,hf^{-1}\,dx^3\wedge dx^4\wedge dx^5\,\,,
\label{cinco}
\eear
where $\epsilon_{(2)}$ is the volume form of the $S^2$. It is not difficult to
verify from eq. (\ref{tres}) that  ${}^*F^{(4)}$ satisfies:
\beq
d\,{}^*F^{(4)}\,=\,-{1\over 2}\,F^{(4)}\wedge F^{(4)}\,,
\label{seis}
\eeq
where the seven-form ${}^*F^{(4)}$ is the Hodge dual of $F^{(4)}$ with respect to
the metric  (\ref{uno}).  Eq.
(\ref{seis}) implies that ${}^*F^{(4)}$ can be represented in terms of a
six-form potential 
$C^{(6)}$ as follows:
\beq
{}^*F^{(4)}\,=\,d\,C^{(6)}\,-\,{1\over 2}\,C^{(3)}\wedge dC^{(3)}\,\,.
\label{siete}
\eeq
By taking the exterior derivative of both sides of (\ref{siete}), one
immediately verifies eq. (\ref{seis}). Moreover, it is not difficult to find the
potential $C^{(6)}$ in our coordinate system. Actually, one can easily check 
that one can take $C^{(6)}$ as:
\bear
C^{(6)}\,&=&\,{1\over 2}\,\sin\varphi\cos\varphi\,f^{-1}\,R^3\,\rho^3\,
dx^0\wedge dx^1\wedge dx^2\wedge d\phi\wedge\epsilon_{(2)}\,-\,\rc\rc
&&-\,{1\over 2}\,{1+h\cos^2\varphi\over \cos\varphi}\,f^{-1}\,
dx^0\wedge dx^1\wedge dx^2\wedge dx^3\wedge dx^4\wedge dx^5\,-\,\rc\rc
&&-\,{1\over 2}\,\sin\varphi\,R^3\,\rho^3\,h\,f^{-1}\,
dx^3\wedge dx^4\wedge dx^5\wedge d\phi\wedge \epsilon_{(2)}\,\,.
\label{ocho}
\eear

\setcounter{equation}{0}
\section{The M5-brane probe}
\medskip
We shall now consider the near-horizon region of the $(M2,M5)$ geometry. In
this region the radial coordinate $r$ is small and one can approximate the
function $f$ appearing in the supergravity solution as 
$f\,\approx\,R^{3}/ r^{3}$. Following the analysis of ref. \cite{gravitons}, we
place an M5-brane probe in this geometry in such a way that it shares three
directions  $(x^3, x^4, x^5)$ with the branes of the background and wraps the
$S^2$ transverse sphere parametrized by the angles $\theta^1$ and $\theta^2$. The
dynamics of the M5-brane probe is determined by its worldvolume action, \ie\ by
the so-called PST action \cite{PST}. In the PST formalism the worldvolume fields
are a three-form field strength $F$ and a scalar field $a$ (the PST scalar). The
action is the sum of three terms:
\beq
S\,=\,T_{M5}\,\int\,d^6\xi\,\Big[\,
{\cal L}_{DBI}\,+\,{\cal L}_{H\tilde H}\,+\,{\cal L}_{WZ}
\,\Big]\,\,,
\label{nueve}
\eeq
where the tension of the M5-brane is $T_{M5}\,=\,1/ (2\pi)^5\,l_p^6$. In the
action (\ref{nueve}) the field strength $F$ is combined with the pullback 
$P[C^{(3)}]$ of the background potential $C^{(3)}$ to form the field $H$:
\beq
H\,=\,F\,-\,P[C^{(3)}]\,\,.
\label{diez}
\eeq
Let us now define the field $\tilde H$ as follows:
\beq
{\tilde H}^{ij}\,=\,{1\over 3!\,\sqrt{-{\rm det}\,g}}\,
{1\over \sqrt{-(\partial a)^2}}\,
\epsilon^{ijklmn}\,\partial_k\,a\,H_{lmn}\,\,,
\label{once}
\eeq
with $g$ being the induced metric on the M5-brane worldvolume. The explicit
form of the three terms of the action is:
\bear
{\cal L}_{DBI}&=& -\sqrt{-{\rm det} (g_{ij}\,+\,\tilde H_{ij})}\,\,,\rc\rc
{\cal L}_{H\tilde H}&=&{1\over 24 (\partial  a)^2}\,\,
\epsilon ^{ijkmnr}\,H_{mnr}\,H_{jkl}\,
g^{ls}\partial_i a \,\partial_s a\,\,,\rc\rc 
{\cal L}_{WZ}&=&{1\over 6!}\epsilon ^{ijklmn}\,\Bigg[\,
P[C^{(6)}]_{ijklmn}\,+\,10\,H_{ijk}\,P[C^{(3)}]_{lmn}\,\Bigg]\,\,.
\label{doce}
\eear
The worldvolume coordinates $\xi^{i}$ ($i=0,\cdots, 5$) will be taken as 
$\xi^{i}=(\,x^0,x^3,x^4,x^5,\theta^1,\theta^2\,)$. In this system of
coordinates the configurations we are interested in are described  by functions
of the type $r=r(t)\,\,$, $\rho=\rho(t)\,\,$ and $\phi=\phi(t)\,\,$, where
$t\equiv x^0$. Moreover, we will assume that the only non-vanishing components
of $H$ are those of $P[C^{(3)}]$, \ie\ $H_{x^3x^4x^5}\equiv H_{345}$ and
$H_{x^0\theta^1 \theta^2}\equiv H_{0*}$. As discussed in ref. \cite{PST}, the
scalar field $a$ is an auxiliary field which, by fixing its gauge symmetry, can
be eliminated from the action at the expense of loosing manifest covariance. In
this paper we will work in the gauge $a=x^0$. In this gauge the only non-zero
component of $\tilde H$ is:
\beq
\tilde H_{\theta^1\theta^2}\,=\,f^{7/6}\,h^{-4/3}\,r^2\rho^2\,
\sqrt{\hat g^{(2)}}\,\,\,H_{345}\,\,,
\label{trece}
\eeq
where $\hat g^{(2)}$ is the determinant of the metric of the two-sphere. 
By using (\ref{trece}) one can easily obtain ${\cal L}_{DBI}$ for our
configurations. Indeed, after a short calculation one gets:
\beq
{\cal L}_{DBI}\,=\,-R^3\rho^2\,\sqrt{\hat g^{(2)}}\,\lambda_1\,
\sqrt{r^{-2}\,f^{-1}\,\,-\,r^{-2}\dot r^2\,-\,
{\dot\rho^2\over 1-\rho^2}\,-\,
(1-\rho^2)\,\dot\phi^2}\,\,,
\label{catorce}
\eeq
where the dot denotes time derivative and $\lambda_1$ is defined as:
\beq
\lambda_1\,\equiv\,\sqrt{h\,f^{-1}\,+\,\big(\,H_{345}\big)^2\,h^{-1}}\,\,.
\label{quince}
\eeq
It is also very easy to prove that the remaining terms of the action are:
\bear
{\cal L}_{H\tilde H}\,+\,{\cal L}_{WZ}&=&{1\over 2}\,F_{345}\,F_{0*}\,-\,
F_{345}\,P[C^{(3)}]_{0*}\,+\,\rc\rc
&&+\,P[C^{(6)}]_{0345*}\,+\,{1\over 2}\,P[C^{(3)}]_{345}\,P[C^{(3)}]_{0*}\,\,,
\label{dseis}
\eear
with $F_{0*}\,\equiv\,F_{x^0\theta^1\theta^2}$ and similarly for the pullbacks
of $C^{(6)}$ and $C^{(3)}$. From eqs. (\ref{cinco}) and (\ref{ocho}) it
follows that:
\bear
P[C^{(6)}]_{0345*}&=& {1\over 2}R^3\rho^3\sin\varphi\,h\,f^{-1}\,
\sqrt{\hat g^{(2)}}\,\dot\phi\,\,,\rc\rc
P[C^{(3)}]_{345}&=&\tan\varphi\,h\,f^{-1}\,\,,\rc\rc
P[C^{(3)}]_{0*}&=&-R^3\rho^3\cos\varphi\,\sqrt{\hat g^{(2)}}\,\dot\phi\,\,.
\label{dsiete}
\eear
By using eq. (\ref{dsiete}) it is straightforward to demonstrate that the sum
of the  last two terms in ${\cal L}_{H\tilde H}\,+\,{\cal L}_{WZ}$ vanishes
and, thus,  we can write:
\beq
{\cal L}_{H\tilde H}\,+\,{\cal L}_{WZ}\,=\,R^3\rho^3\,F_{345}\cos\varphi\,
\sqrt{\hat g^{(2)}}\,\dot\phi\,+\,{1\over 2}\,F_{345}\,F_{0*}\,\,.
\label{docho}
\eeq
Let us assume that $F_{0*}\,=\,\sqrt{\hat g^{(2)}}\,f_{0*}$ with $f_{0*}$
independent of the angles of the $S^2$. With this ansatz for the electric
component of $F$, the action can be written as:
\beq
S\,=\,\int dt\,dx^3\,dx^4\,dx^5\,\,{\cal L}\,\,,
\label{dnueve}
\eeq
with the lagrangian density ${\cal L}$ given by:
\bear
{\cal L}&=&4\pi R^3\,T_{M5}\,\Bigg[\,-\rho^2\,
\lambda_1\, \sqrt{r^{-2}\,f^{-1}\,\,-\,r^{-2}\dot r^2\,-\,
{\dot\rho^2\over 1-\rho^2}\,-\, (1-\rho^2)\,\dot\phi^2}\,+\,\rc\rc
&&+\,\lambda_2\,\rho^3\dot\phi\,+\,{1\over 2 R^3}\,
F_{345}\,f_{0*}\,\Bigg]\,\,.
\label{veinte}
\eear
In eq. (\ref{veinte}) we have defined $\lambda_2$ as:
\beq
\lambda_2\,\equiv\,F_{345}\,\cos\varphi\,\,.
\label{vuno}
\eeq
As in ref. \cite{gravitons}, it is interesting to characterize the spreading of
the M5-brane in the $x^3x^4x^5$ directions by means of the flux of the
worldvolume gauge field $F$. We shall parametrize this flux as follows:
\beq
\int dx^3\,dx^4\,dx^5\,\,F\,=\,{2\pi\over T_{M2}}\,\,N'\,\,,
\label{vdos}
\eeq
where $T_{M2}\,=\,1/(2\pi)^2\,l_{p}^3$ in the tension of the M2-brane. Notice
that, when the coordinates $x^3x^4x^5$ are compact, the condition (\ref{vdos})
is just the M-theory flux quantization condition found in ref. \cite{flux}, with
the flux number $N'$ being an integer for topological reasons. 

In order to perform a canonical hamiltonian analysis of this system, let us
introduce the density of momenta:
\bear
{\cal P}_r&=&{\partial {\cal L}\over \partial \dot r}\,\equiv\,
4\pi R^{3}\,T_{M5}\,\lambda_1\,\pi_r\,\,,\rc\rc
{\cal P}_{\rho}&=&{\partial {\cal L}\over \partial \dot \rho}\,\equiv\,
4\pi R^{3}\,T_{M5}\,\lambda_1\,\pi_{\rho}\,\,,\rc\rc
{\cal P}_{\phi}&=&{\partial {\cal L}\over \partial \dot \phi}\,\equiv\,
4\pi R^{3}\,T_{M5}\,\lambda_1\,\pi_{\phi}\,\,,
\label{vtres}
\eear
where we have defined the reduced momenta $\pi_r$, $\pi_{\rho}$ and
$\pi_{\phi}$.  From the explicit value of ${\cal L}$ (eq. (\ref{veinte})), we
get:
\bear
 \pi_r&=&{\rho^{2}\over r^2}\,\,{\dot r\over
\sqrt{r^{-2}\,f^{-1}\,\,-\,r^{-2}\dot r^2\,-\,
{\dot\rho^2\over 1-\rho^2}\,-\,(1-\rho^2)\,\dot\phi^2}}\,\,,\rc\rc
\pi_{\rho}&=&{\rho^{2}\over 1- \rho^2}\,\,{\dot \rho\over
\sqrt{r^{-2}\,f^{-1}\,\,-\,r^{-2}\dot r^2\,-\,
{\dot\rho^2\over 1-\rho^2}\,-\,(1-\rho^2)\,\dot\phi^2}}\,\,,\rc\rc
 \pi_{\phi}&=&(1-\rho^2)\rho^{2}\,\,{\dot \phi\over
\sqrt{r^{-2}\,f^{-1}\,\,-\,r^{-2}\dot r^2\,-\,
{\dot\rho^2\over 1-\rho^2}\,-\,(1-\rho^2)\,\dot\phi^2}}
\,\,+\,\,\Lambda\,\rho^{3}\,\,,
\label{vcuatro}
\eear
where we have introduced the quantity  
$\Lambda \equiv \lambda_2 /\lambda_1$. The hamiltonian density of the system is:
\beq
{\cal H}\,=\,\dot r\,{\cal P}_r\,+\,\dot\rho\,{\cal P}_{\rho}\,
+\,\dot\phi{\cal P}_{\phi}\,
+\,F_{0*}\,{\partial {\cal L}\over \partial F_{0*}} \,-\,{\cal L}\,\,.
\label{vcinco}
\eeq
After a short calculation one can prove that ${\cal H}$ is given by:
\beq
{\cal H}\,=\,4\pi R^3\,T_{M5}\,\lambda_1\,
r^{-1}\,f^{-{1\over 2}}\,\Bigg[\,r^2\,\pi_r^2\,+\,\rho^{4}\,+\,
(1-\rho^2)\,\pi_{\rho}^2\,+\,
{\Big(\pi_{\phi}-\Lambda\rho^{3}\Big)^2\over 1-\rho^2}
\,\,\Bigg]^{{1\over 2}}\,\,.
\label{vseis}
\eeq

\setcounter{equation}{0}
\section{Giant graviton configurations}
\medskip
By inspecting the line element displayed in eq. (\ref{cuatro}) one easily
concludes that the coordinate $\rho$ plays the role of the size of the system
on the $S^2$ sphere. We are interested in finding configurations of fixed size,
\ie\ those solutions of the equations of motion with constant $\rho$. By
comparing the hamiltonian density written in (\ref{vseis}) with the one studied
in ref. \cite{gravitons}, it is not difficult to realize that these fixed size
solutions exist if the quantity $\Lambda$ takes the value $\Lambda=1$. Indeed,
if this condition holds, the hamiltonian density ${\cal H}$ can be put as:
\beq
{\cal H}\,=\,4\pi R^3\,T_{M5}\,\lambda_1\,
r^{-1}\,f^{-{1\over 2}}\,\Bigg[\,\pi_{\phi}^2\,+\,r^2\,\pi_r^2\,+\,
(1-\rho^2)\,\pi_{\rho}^2\,+\,
{\Big(\pi_{\phi}\rho-\rho^{2}\Big)^2\over 1-\rho^2}
\,\,\Bigg]^{{1\over 2}}\,\,,
\label{vsiete}
\eeq
and, as we will verify soon, one can easily find constant $\rho$ solutions of the
equations of motion for the hamiltonian (\ref{vsiete}). Moreover, by using the
value of $P[C^{(3)}]_{345}$ given in eq. (\ref{dsiete}), one can write
$\lambda_1$ as:
\beq
\lambda_1^2\,=\,\cos^2\varphi\,F_{345}^2\,+\,f^{-1}\,
\Big(\,F_{345}\sin\varphi\,-\,{1\over \cos\varphi}\Big)^{2}\,\,.
\label{vocho}
\eeq
Taking into account the definition of $\lambda_2$ (eq. (\ref{vuno})), it
follows that the condition  $\Lambda\,=\,1$ (or $\lambda_1=\lambda_2$) is
equivalent to have the following constant value of the worldvolume gauge field:
\beq
F_{345}\,=\,{1\over \sin\varphi\cos\varphi}\,=\,2\csc (2\varphi)\,\,.
\label{vnueve}
\eeq
It follows  from eq. (\ref{vcuatro}) that for a configuration with $\dot\rho=0$
the momentum $\pi_{\rho}$ necessarily vanishes and, in particular, one must
require that $\dot\pi_{\rho}=0$. Then, the hamiltonian equations of motion
imply that $\partial {\cal H}/\partial \rho$ must be zero, which happens if the
last term inside the square root of the right-hand side of eq. (\ref{vsiete})
vanishes, \ie\ when $\pi_{\phi}\rho\,-\,\rho^{2}=0$. This occurs either when
$\rho=0$ or else when the angular momentum  $\pi_{\phi}$ is:
\beq
\pi_{\phi}\,=\,\rho\,\,.
\label{treinta}
\eeq
In order to clarify the nature of these solutions, let us invert the relation
between $\pi_{\phi}$ and $\dot\phi$ (eq. (\ref{vcuatro})). After a simple
calculation one gets:
\beq
\dot\phi\,=\,
{\pi_{\phi}-\rho^{3}\over 1-\rho^2}\,\,
{\Bigg[\,r^{-2}\big(f^{-1}\,-\,\dot r^2\,\big)\,-\,
{\dot\rho^2\over 1-\rho^2}\,\Bigg]^{{1\over 2}}\over
\Bigg[\,\pi_{\phi}^2\,+\,
{\big(\,\pi_{\phi}\rho\,-\,\rho^{2}\,\big)^2
\over 1-\rho^2}\,\Bigg]^{{1\over 2}}}\,\,.
\label{tuno}
\eeq
By taking $\dot\rho=\pi_{\phi}\rho\,-\,\rho^{2}=0$ on the right-hand side of
eq. (\ref{tuno}), one finds the following relation between $\dot\phi$ and
$\dot r\,$:
\beq
f\,\big[\,r^2\,\dot\phi^2\,+\,\dot r^2\,]\,=\,1\,\,.
\label{tdos}
\eeq
Remarkably, eq. (\ref{tdos}) is the condition satisfied by a particle which
moves in the $(r,\phi)$ plane at $\rho=0$ along a null trajectory (\ie\ with 
$ds^2=0$) in the metric (\ref{uno}). Therefore, our brane probe configurations
behave as a massless particle: the so-called giant graviton. The point $\rho=0$
can be interpreted as the ``center of mass" of the expanded brane. Actually,
if one defines the velocity vector  ${\bf v}$ as 
${\bf v}\,=\,(v^{\underline{r}}, v^{\underline{\phi}})\,\equiv\,
f^{{1\over 2}}\,(\dot r\,,\,r\dot\phi) $, eq. (\ref{tdos}) is equivalent to
the condition $(v^{\underline{r}})^2\,+\,(v^{\underline{\phi}})^2\,=\,1$ and,
thus, the center of mass of the giant graviton moves at the speed of light. On
the other hand,  the angular momentum density  ${\cal P}_{\phi}$ for the
$\pi_{\phi}=\rho$ solution can be obtained from eq. (\ref{vtres}), namely:
\beq
{\cal P}_{\phi}\,=\,{T_{M2}\over 2\pi}\,F_{345}\,N\,\rho\,\,.
\label{ttres}
\eeq
Moreover, by integrating the densities ${\cal P}_{\phi}$ and 
${\cal P}_{r}$ along the $x^{3}x^4x^5$ directions, one gets the values of the
momenta $p_\phi$ and $p_r$:
\beq
p_\phi\,=\,\int dx^{3}\,dx^4\,\,dx^5\,{\cal P}_{\phi}\,\,,
\,\,\,\,\,\,\,\,\,\,\,\,\,\,\,\,\,\,\,\,\,\,
p_r\,=\,\int dx^{3}\,dx^4\,\,dx^5\,\,{\cal P}_{r}\,\,.
\label{tcuatro}
\eeq
By using the value of the momentum density ${\cal P}_{\phi}$ displayed in eq. 
(\ref{ttres}), together with the flux quantization condition (\ref{vdos}), one
gets the following value of $p_\phi$:
\beq
p_\phi\,=\,N\,N'\,\rho\,\,,
\label{tcinco}
\eeq
which implies that the size $\rho$ of the wrapped brane increases with its
angular momentum $p_\phi$. As $0\le\rho\le 1$, the momentum $p_\phi$ has a
maximum given by  $p_\phi^{max}\,=\,N\,N'$. This maximum is reached when
$\rho=1$ and its existence is a manifestation of the stringy exclusion
principle. 

In order to analyze the energy of the giant graviton solution, let 
${\cal G}_{MN}$ be the metric elements of eqs. (\ref{uno}) and (\ref{cuatro})
at the point $\rho=0$. Then, it is straightforward to verify that the
hamiltonian $H_{GG}$ of the giant graviton configurations is:
\beq
 H_{GG}\,=\,\sqrt{-{\cal G}_{tt}}\,\Bigg[\,
 {p_{\phi}^2\over {\cal G}_{\phi\phi}}\,+\,
{ p_{r}^2\over {\cal G}_{rr}}\,\,\Bigg]^{{1\over 2}}\,\,,
\label{tseis}
\eeq
which is exactly the one corresponding to a massless particle which moves in
the $(r,\phi)$ plane under the action of the metric ${\cal G}_{MN}$. By
substituting in eq. (\ref{tseis}) the explicit values of the ${\cal G}_{MN}$'s,
one can write 
$H_{GG}$ as:
\beq
 H_{GG}\,=\,R^{-{3\over 2}}\,\,
\Bigg[\,r^{3}\, p_{r}^2\,+\,r\, p_{\phi}^2
\,\Bigg]^{{1\over 2}}\,\,.
\label{tsiete}
\eeq
By using the conservation of energy, one can integrate the equations of motion
and get the functions $r(t)$ and $\phi(t)$. It turns out that the corresponding
equations coincide with one of the cases studied in ref. \cite{gravitons}.
Therefore, we simply write the results of this integration and refer to
\cite{gravitons} for the details of the calculation. One gets:
\bear
r\,&=&{r_{*}\over 1\,+\,{r_{*}\over 4R^3}\,(t\,-\,t_{*})^2}\,\,,\rc\rc\rc
\tan\Big[{\phi\,-\,\phi_*\over 2}\Big]&=&{1\over 2R}\,\,
\Big({r_{*}\over R}\Big)^{{1\over 2}}\,(t\,-\,t_{*})\,\,,
\label{tocho}
\eear
where $r_{*}$, $\phi_*$ and $t_{*}$ are constants. Notice that $r\le r_{*}$ and
that $r\rightarrow 0$ as $t\rightarrow\infty$, which means that the giant
graviton always falls asymptotically to the center of the potential. 

It is also interesting to study the volume occupied by the M5-brane probe along
the $x^{3}x^4x^5$ directions. By plugging the value (\ref{vnueve}) of the
worldvolume gauge field into the flux quantization condition (\ref{vdos}), one
gets that this volume is:
\beq
\int dx^{3}\,dx^4\,\,dx^5\,=\,{\pi N'\over T_{M2}}\,\,\,\sin(2\varphi)\,\,.
\label{tnueve}
\eeq
When $\varphi\rightarrow 0$, the M2-brane component of the background bound
state disappears and we are left with a M5-brane background. For fixed $N'$, it
follows from eq. (\ref{tnueve}) that the three directions of the M5-brane probe
which are parallel to the background collapse and, therefore, the M5-brane
probe is effectively converted into a M2-brane, in agreement  with the results
of ref. \cite{GST}. 

The gauge field $F$ of the PST action satisfies a generalized self-duality
condition which relates its electric and magnetic components. In order to get
this self-duality constraint  one must  use both the equations of motion and the
symmetries of the  PST action \cite{PST}. In our case, this condition
reduces to:
\beq
{\partial {\cal L}\over \partial F_{345}}\,=\,0\,\,.
\label{cuarenta}
\eeq
Indeed, after using the explicit expression of ${\cal L}$ (eq. (\ref{veinte})),
and solving eq. (\ref{cuarenta}) for $f_{0*}$, one gets:
\beq
f_{0*}\,=\,2R^3\Bigg[\,\rho^2\,{H_{345}\over \lambda_1\,h}\,
\sqrt{r^{-2}\,f^{-1}\,\,-\,r^{-2}\dot r^2\,-\,
{\dot\rho^2\over 1-\rho^2}\,-\,(1-\rho^2)\,\dot\phi^2}\,-\,
\cos\varphi\,\,\rho^3\,\dot\phi\,\Bigg]\,\,.
\label{cuno}
\eeq
By substituting on the right-hand side of eq. (\ref{cuno}) the values
corresponding to our giant graviton configurations, one gets a vanishing
result, \ie:
\beq
{f_{0*}}_{\big|{GG}}\,=\,0\,\,.
\label{cdos}
\eeq
Thus, our expanded graviton solutions have zero electric  
field on the M5-brane worldvolume. 

\setcounter{equation}{0}
\section{Supersymmetry}
\medskip
Let us now examine the supersymmetry of our configurations. First of all, we
consider the supersymmetry preserved by the background. As the solution of
D=11 supergravity we are dealing with is purely bosonic, it is only invariant
under those supersymmetry transformations which do no change the gravitino field 
$\psi_M$. This field transforms as:
\beq
\delta\psi_M\,=\,D_M\,\epsilon\,+\,{1\over 288}\,
\Bigg(\,\Gamma_{M}^{\,\,N_1\cdots N_4}\,-\,8\delta_{M}^{N_1}\,
\Gamma^{\,\,N_2\cdots N_4}\,\Bigg)\,\epsilon\,\,
F^{(4)}_{N_1\cdots N_4}\,\,.
\label{ctres}
\eeq
The spinors $\epsilon$ for which the right-hand side of eq. (\ref{ctres}) vanish
are the Killing spinors of the background. It is not difficult to find them in
our case. Actually, if we define the matrix
$\Upsilon\,=\,\Gamma_{\underline{\phi}}\,\Gamma_{*}$ 
with $\Gamma_{*}\,\equiv\,\Gamma_{\underline{\theta^1\theta^2}}$, they
can be parametrized as follows: 
\beq
\epsilon\,=\, e^{{\alpha\over 2}\,\Gamma_{\underline{x^{3}x^{4}x^{5}}}}\,\,
e^{-{\beta\over 2}\,\,\Upsilon}\,\,
\,\,\,\hat\epsilon\,\,,
\label{ccuatro}
\eeq
where $\alpha$ and $\beta$ are:
\bear
\sin\alpha&=&\,f^{-{1\over 2}}\,h^{{1\over 2}}\sin\varphi\,\,,
\,\,\,\,\,\,\,\,\,\,\,\,\,\,\,\,\,\,\,\,\,\,
\cos\alpha\,=\,h^{{1\over 2}}\cos\varphi\,\,,\rc\rc
\sin\beta&=&\rho\,\,,
\,\,\,\,\,\,\,\,\,\,\,\,\,\,\,\,\,\,\,\,\,\,
\,\,\,\,\,\,\,\,\,\,\,\,\,\,\,\,\,\,\,\,\,\,\,\,\,\,\,\,\,\,
\cos\beta\,=\,\sqrt{1-\rho^2}\,\,,
\label{ccinco}
\eear
and $\hat\epsilon$ is independent of $\rho$ and satisfies:
\beq
\Gamma_{\underline{x^{0}\cdots x^{5}}}\,\,\hat\epsilon\,=\,\hat\epsilon\,\,.
\label{cseis}
\eeq
By working out the condition $\delta\psi_M\,=\,0$ one can determine
$\hat\epsilon$ completely. We will not reproduce this calculation here since
the representation (\ref{ccuatro}) is enough for our purposes. Let us however
mention that that it follows from this analysis that the (M2,M5) background is
$1/2$ supersymmetric. 

The number of supersymmetries preserved by the M5-brane probe is the number
independent solutions of the equation $\Gamma_{\kappa}\epsilon=\epsilon$, where 
$\epsilon$ is one of the Killing spinors (\ref{ccuatro}) and 
$\Gamma_{\kappa}$ is the $\kappa$-symmetry matrix of the PST
formalism \cite{PST, APPS}. In order to write the expression of this matrix, let
us define the following quantities:
\beq
\nu_p\,\equiv {\partial_p a\over \sqrt{-(\partial a)^2}},
\,\,\,\,\,\,\,\,\,\,\,\,\,\,\,\,\,\,\,\,
\,\,\,\,\,\,\,\,\,\,\,\,\,\,\,\,\,\,\,\,\,\,
t^m\,\equiv\,{1\over 8}\,
\epsilon^{mn_1n_2p_1p_2q}\,\tilde H_{n_1n_2}\,\tilde H_{p_1p_2}\,\nu_q\,\,.
\label{csiete}
\eeq
Then, the  $\kappa$-symmetry matrix is:
\beq
\Gamma_{\kappa}=-{\nu_m\gamma^m\over \sqrt{-{\rm det} (g+\tilde H)}}\,\,
\Bigg[\,\gamma_n t^n\,+\,
{\sqrt{-g}\over 2}\,\gamma^{np}\,\tilde H_{np}\,
+\,{1\over 5!}\gamma_{i_1\cdots i_5}\,\epsilon^{i_1\cdots i_5n}\nu_n\,\,
\Bigg]\,\,.
\label{cocho}
\eeq
In eq. (\ref{cocho}) $\gamma_{i_1i_2\cdots }$  are antisymmetrized
products of the worldvolume Dirac matrices \break\hfill
$\gamma_i=\partial_iX^M\,E^{\underline M}_M\,
\Gamma_{\underline M}$. In our case
the vector $t^m$ is zero and the only non-zero component of $\nu_m$ is:
 $\nu_0\,=\,\sqrt{-G_{tt}}$. Using these facts, after some calculation, one
can represent $\Gamma_{\kappa}$ as:
\bear
\Gamma_{\kappa}&=&{\,\,1\over
\sqrt{-G_{tt}\,-\,G_{\phi\phi}\dot\phi^2\,-\,G_{rr}\dot r^2}}\,\,\times\rc\rc
&&\times\Bigg[\,\sqrt{-G_{tt}}\,\Gamma_{\underline{x^0}}\,+\,
\dot\phi\sqrt{G_{\phi\phi}}\,\Gamma_{\underline{\phi}}\,+\,
\dot r\sqrt{G_{rr}}\,\Gamma_{\underline{r}}\,\Bigg]\,\Gamma_{*}\,\,
e^{-\eta\,\Gamma_{\underline{x^{3}x^{4}x^{5}}}}\,\,,
\label{cnueve}
\eear
with $\eta$ given by:
\beq
\sin\eta\,=\,{f^{-{1\over 2}}\,h^{{1\over 2}}\over \lambda_1}\,\,,
\,\,\,\,\,\,\,\,\,\,\,\,\,\,\,\,\,\,\,\,\,\,
\cos\eta\,=\,{H_{345}\,h^{-{1\over 2}}\over \lambda_1}\,\,. 
\label{cincuenta}
\eeq
By using eqs. (\ref{cnueve}) and (\ref{ccuatro}), the equation 
$\Gamma_{\kappa}\epsilon=\epsilon$ takes the form:
\bear
{1\over
\sqrt{-G_{tt}\,-\,G_{\phi\phi}\dot\phi^2\,-\,G_{rr}\dot r^2}}
&&\Bigg[\,\sqrt{-G_{tt}}\,\Gamma_{\underline{x^0}}+
\dot\phi\sqrt{G_{\phi\phi}}\,\Gamma_{\underline{\phi}}+
\dot r\sqrt{G_{rr}}\,\Gamma_{\underline{r}}\,\Bigg]\,\Gamma_{*}\,\,
e^{-{\beta\over 2}\,\Upsilon}\hat\epsilon\,=\rc\rc
&&=\,e^{(\alpha-\eta)\,\,
\Gamma_{\underline{x^{3}x^{4}x^{5}}}}\,
e^{-{\beta\over 2}\,\Upsilon}\hat\epsilon\,\,.
\label{ciuno}
\eear
Let us now evaluate eq. (\ref{ciuno}) for our solution. First of all, one can
verify that, when the worldvolume gauge field $F_{345}$ takes the value 
(\ref{vnueve}), the angles $\alpha$ and $\eta$ are equal and, thus, the
dependence on $\Gamma_{\underline{x^{3}x^{4}x^{5}}}$ of the right-hand side of 
(\ref{ciuno}) disappears. Moreover, using the condition (\ref{tdos}), and
performing some simple manipulations, one can convert eq.  (\ref{ciuno}) into:
\beq
\Bigg[\,f^{-{1\over 4}}\,e^{\beta\,\Upsilon}\,
\Gamma_{\underline{x^0\phi}}-
\dot\phi\,r\,f^{{1\over 4}}\,\sqrt{1-\rho^2}+
\dot r f^{{1\over 4}}\,e^{\beta\,\Upsilon}\,
\Gamma_{\underline{r\phi}}\,\Bigg]\,\,
\hat\epsilon\,=\,\rho\, r\, f^{{1\over 4}}\,\dot\phi\,\Upsilon
\,\,\hat\epsilon\,\,.
\label{cidos}
\eeq
If, in particular, we take $\rho=0$ in eq. (\ref{cidos}), one arrives at:
\beq
\Bigg[\,f^{-{1\over 4}}\,\Gamma_{\underline{x^0\phi}}
-\dot\phi\,r\,f^{{1\over 4}}+
\dot r f^{{1\over 4}}\Gamma_{\underline{r\phi}}\,\Bigg]\,\,
\hat\epsilon\,=\,0\,\,.
\label{citres}
\eeq
Remarkably, if eq. (\ref{citres}) holds, then eq. (\ref{cidos}) is satisfied for
an arbitrary value of $\rho$. Thus, eq. (\ref{citres}) is equivalent to the
$\kappa$-symmetry condition $\Gamma_{\kappa}\epsilon=\epsilon$.  In order to
interpret (\ref{citres}), let us define the matrix 
$\Gamma_v\,\equiv\,v^{\underline{r}}\,\Gamma_{\underline{r}}\,+\,
v^{\underline{\phi}}\,\Gamma_{\underline{\phi}}$, where $v^{\underline{r}}$ and 
$v^{\underline{\phi}}$ are the components of the center of mass velocity vector
${\bf v}$ defined above. This matrix is such that 
$(\,\Gamma_v\,)^2\,=\,1$, and one can prove that eq. (\ref{citres}) can
be written as:
\beq
\Gamma_{\underline{x^0}}\,\Gamma_v\,\hat\epsilon\,\,=\,\,
\hat\epsilon\,\,.
\label{cicuatro}
\eeq
Taking into account the relation (\ref{ccuatro}) between $\hat\epsilon$ and
$\epsilon$, and using the fact that $\Gamma_{\underline{x^0}}\,\Gamma_v$
commutes with $\Gamma_{\underline{x^{3}x^{4}x^{5}}}$, one can recast eq. 
(\ref{cicuatro}) as:
\beq
\Gamma_{\underline{x^0}}\,\Gamma_v\,\epsilon_{{\,\big |}_{\rho=0}}\,\,=\,\,
\epsilon_{{\,\big |}_{\rho=0}}\,\,,
\label{cicinco}
\eeq
which is the supersymmetry projection induced by a massless particle 
moving in the direction of ${\bf v}$ at $\rho=0$. Notice that, however,  the
background projector 
$\Gamma_{\underline{x^{0}\cdots x^{5}}}$ does not commute with 
$\Gamma_{\underline{x^0}}\,\Gamma_v$ and, therefore, eq. (\ref{cseis}) and
(\ref{cicuatro})  cannot be imposed at the same time. Thus, the M5-brane probe
breaks completely the supersymmetry of the background. The interesting point in
this result is that this supersymmetry breaking is just identical to the one
corresponding to a massless particle, which constitutes a confirmation of our
interpretation of the giant graviton configurations. 

\setcounter{equation}{0}
\section{Summary and conclusions}
\medskip
In this paper we have found giant graviton configurations of an M5-brane probe
in the D=11 supergravity background created by a stack of (M2,M5) bound
states. We have solved the probe equations of motion and we have checked that
the corresponding solution behaves as an expanded massless particle propagating
in the (M2,M5) background. We have also checked that the probe breaks the
supersymmetry of the background exactly in the same  way as a massless particle
moving along the  trajectory of the center of mass of the probe. Our
results generalize those of refs. \cite{GST}-\cite{gravitons} and,
hopefully, could be useful to shed light on the nature of the blown up graviton
systems.

\medskip
\section{ Acknowledgments}
\medskip
We are grateful to J. M. Sanchez de Santos for discussions. 
This work was
supported in part by DGICYT under grant PB96-0960,  by CICYT under
grant  AEN99-0589-CO2 and by Xunta de Galicia under  grant
PGIDT00-PXI-20609.

\end{document}